\documentclass[pra,twocolumn,showpacs,preprintnumbers,amsmath,amssymb,floatfix]{revtex4}
\usepackage{epsfig}

\newcommand{\beq}{\begin{equation}}
\newcommand{\eeq}{\end{equation}}
\newcommand{\beqa}{\begin{eqnarray}}
\newcommand{\eeqa}{\end{eqnarray}}
\newcommand{\ba}{\begin{array}}
\newcommand{\ea}{\end{array}} 

\begin{document} 



\title{Expansion of a Fermi Cloud in the BCS-BEC Crossover} 
\author{G. Diana$^{1}$, N. Manini$^{1}$, and L. Salasnich$^{1,2}$} 
\affiliation{$^1$Dipartimento di Fisica and CNR-INFM, Universit\`a di Milano,
Via Celoria 16, 20133 Milano, Italy
\\ 
$^2$CNR-INFM and CNISM, Unit\`a di Milano, Via Celoria 16, 20133 Milano} 

\begin{abstract} 
We study the free expansion of a dilute
two-component Fermi gas with attractive 
interspecies interaction in the BCS-BEC crossover. 
We apply a time-dependent parameter-free density-functional theory 
by using two choices of the equation of state:  
an analytic formula based on Monte Carlo data 
and the mean-field equation of state resulting from the 
extended BCS equations. The calculated axial and transverse radii 
and the aspect ratio of the expanding cloud are compared 
to experimental data on vapors of $^6$Li atoms. 
Remarkably, the mean-field theory shows a better agreement with the
experiments than the theory based on the Monte Carlo equation of state.
Both theories predict a measurable dependence of the aspect ratio on
expansion time and on scattering length.
\end{abstract} 


\maketitle



\section{Introduction} 

Current experiments with cold vapors of $^{6}$Li and $^{40}$K atoms 
can operate in the regime of deep Fermi degeneracy. 
The available experimental data on two-hyperfine-component Fermi gases are
concentrated across a Feshbach resonance, where the s-wave scattering
length $a_F$ of the interatomic potential varies from large negative to
large positive values \cite{ohara,greiner,jochim,bourdel} and where a
crossover from a Bardeen-Cooper-Schrieffer (BCS) superfluid to a
Bose-Einstein condensate (BEC) of molecular pairs has been predicted
\cite{leggett,nozieres,engelbrecht}.
In these experiments, the Fermi cloud is dilute because the effective range
$R_0$ of the interaction is much smaller than the mean interparticle
distance, i.e. $k_F R_0 \ll 1$ where $k_F=(3\pi^2 n)^{1/3}$ is the Fermi
wave vector and $n$ is the gas number density.
Even in this dilute regime the s-wave scattering length $a_F$ can be made
very large: the interaction parameter $k_F a_F$ diverges and changes sign
at a Feshbach resonance, despite $k_F R_0$ remaining small
\cite{ohara,greiner,jochim,pitaevskii}.

Recent experimental and theoretical investigations studied the density
profiles \cite{kinast,bartenstein,perali}, collective excitations
\cite{kinast,bartenstein,stringari,combescot,minguzzi,heiselberg,kim,manini},
condensate fraction \cite{zwierlein,ortiz,salasnich,giorgini} and vortices
\cite{bulgac,ketterle} of the fermion cloud through the BCS-BEC crossover.
In this letter we analyze the free expansion of the Fermi gas through this
crossover by using a parameter-free time-dependent density-functional
theory \cite{kim,manini} based on the bulk equation of state of the
superfluid, and including a quantum-pressure term.
We adopt two possible equations of state: a reliable analytical
interpolating formula based on bulk Monte Carlo results \cite{manini} and
the mean-field equation of state based on extendend BCS equations
\cite{leggett,nozieres,engelbrecht,minguzzi}.
Experimentally, the free expansion of superfluid $^6$Li clouds was observed
by O'Hara {\it et al.}\ \cite{ohara} and by Bourdel {\it et al.}\
\cite{bourdel}.
The comparison of our theory with these experimental data shows that the
effects of interaction could be 
detected during the
expansion if the thermal component was negligible.
In addition, by using local scaling equations, we investigate the long-time
dynamics of the Fermi gas predicting novel and 
measurable
effects of interaction on the time evolution of the expansion process.

\section{Theory} 

To describe the dynamics of a zero-temperature Fermi cloud in the external
potential $U({\bf r},t)$ we use a hydrodynamic model with a von
Weizs\"acker quantum-pressure term.
This theoretical approach is expected to be reliable for studying the
collective dynamics of the Fermi gas \cite{kim,manini}. The action
functional $A[\psi]$ of the theory depends on the superfluid order
parameter $\psi({\bf r},t)$ as follows
\beq 
A = \int dt \; d^3{\bf r} \; 
{\cal L}(\psi , \partial_t\psi , \nabla \psi) \; , 
\eeq 
where the Lagrangian density reads 
\beq 
{\cal L} = i\hbar \; \psi^* \partial_t \psi + {\hbar^2 \over 2m} 
\psi^* \nabla^2 \psi - U |\psi|^2 - {\cal E}(|\psi|^2)|\psi|^2 
. 
\eeq
${\cal E}$ represents the bulk energy per particle of the system, which is
conveniently expressed as a function of the number density $n=|\psi|^2$ by
the following equation:
\beq 
{\cal E}(n) = {3\over 5} \; 
{\hbar^2 k_F^2\over 2m} \; f(y) \; , 
\eeq 
where
$f(y)$ is a function of the inverse interaction parameter 
$y=(k_F a_F)^{-1}$. In the weakly attractive regime ($y\ll -1$) one expects 
a BCS Fermi gas of weakly bound Cooper pairs where the 
superfluid gap energy $\Delta$ is exponentially small. 
In the so-called unitarity limit ($y=0$) one expects that
the energy per particle is proportional to that 
of a non-interacting Fermi gas with a $n$-independent coefficient 
$f(0)=0.42$ \cite{baker}. 
In the weak-coupling BEC regime ($y\gg 1$), 
a weakly repulsive Bose gas of dimers 
of mass $m_B=2m$ and density $n_B=n/2$ is expected.  
Such Bose-condensed molecules interact with a positive scattering length
$a_B=0.6 a_F$ \cite{petrov,montecarlo}.
%
The function $f(y)$ is modelled by the analytical formula 
\beq 
f(y) = \alpha_1 - \alpha_2 
\arctan{\left( \alpha_3 \; y \; 
{\beta_1 + |y| \over \beta_2 + |y|} \right)}    
\eeq 
recently derived \cite{manini} from Monte Carlo (MC) simulations 
\cite{montecarlo,carlson} and asymptotic expressions. 
Table 1 of Ref.~\cite{manini} reports the values of 
the interpolating $\alpha_1$, $\alpha_2$, $\alpha_3$, $\beta_1$,
and $\beta_2$.

In the present investigation we take an axially symmmetric harmonic
potential as confining trap
\beq 
U({\bf r},t) = {m \over 2} 
\left[ {\bar \omega}_{\rho}(t)^2 (x^2 + y^2) + 
{\bar \omega}_z(t)^2 z^2 \right]
,  
\eeq 
where ${\bar \omega}_j(t)=\omega_j\Theta(-t)$, with 
$j=1,2,3={\rho},{\rho},z$ and $\Theta(t)$ is the step function, 
so that, after the external trap is switched off at $t>0$, 
the Fermi cloud performs a free expansion. 
The Euler-Lagrange equation for the field $\psi({\bf r},t)$ 
is obtained by minimizing the action functional 
of Eqs.~(1,2). This leads to a time-dependent 
nonlinear Schr\"odinger equation (TDNLSE):  
\beq 
i\hbar \; \partial_t \psi = \left[ -{\hbar^2 \over 2m} \nabla^2 
+ U + \mu(|\psi|^2) \right] \psi  \; .    
\eeq
The nonlinear term $\mu$ is the bulk chemical potential 
of the system. Like the energy $\cal E$ of Eq.~(3), 
also the bulk chemical 
potential $\mu$ is a function of the number density $n$. 
The MC chemical potential is related to the MC energy 
by the thermodynamical formula 
\beq 
\mu (n) = {\partial \left( n {\cal E}( n ) \right) 
\over \partial n } = {\hbar^2 k_F^2\over 2m} 
\left[ f(y) - {y \over 5} f'(y) \right]
. 
\eeq

Instead of the MC equation of state (7) based on (4), 
in the TDNLSE (6) one can plug the mean-field equation 
of state, obtained from the extendend BCS (EBCS) equations 
\cite{leggett,nozieres,engelbrecht}. 
In this scheme, the chemical 
potential $\mu$ and the gap energy $\Delta$ of the uniform Fermi 
gas are found by solving the following EBCS equations
\beqa 
-{1\over a_F} &=& {2 (2m)^{1/2} \over \pi \hbar^3} \,
\Delta^{1/2} \, I_1\!\left({\mu \over \Delta}\right)
, 
\label{gbcs1} 
\\
n &=& {(2m)^{3/2} \over 2 \pi^2 \hbar^3} \,
\Delta^{3/2} \, I_2\!\left({\mu \over \Delta}\right)
,
\label{gbcs2} 
\eeqa
where $I_1(x)$ and $I_2(x)$ are two monotonic
functions which can be expressed in terms of elliptic 
integrals \cite{salasnich,marini}. By solving these two EBCS equations 
we obtain the chemical potential $\mu$ as a function of $n$ and $a_F$, 
which can be inserted into the TDNLSE (6). 

\section{Expansion of Fermi gas and scaling equations} 

\begin{figure}
\epsfig{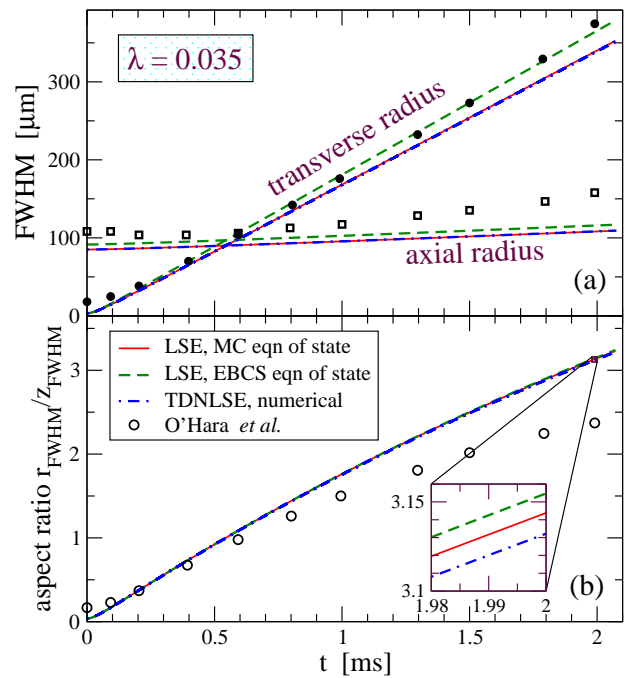}
\caption{\label{t-radii}
(Color online). Expansion of a cloud of 
$1.5\cdot 10^5$ $^6$Li atoms released from a trap as realized 
in Ref.~\protect\cite{ohara}, with anisotropy 
$\lambda=\omega_z/\omega_{\rho}=0.035$. 
(a) Transverse and axial radii of the $^6$Li atomic cloud 
close to the unitarity limit: $a_F=-0.14 a_z$ which 
corresponds to $y=-0.16$. 
(b) Aspect ratio of the cloud as function of the 
time $t$.
Circles and squares: experimental data of
Ref.~\protect\cite{ohara};
dot-dashed lines: TDNLSE with the MC equation of state; 
solid lines: LSE with the MC equation of state; 
dashed lines: LSE with the EBCS equation of state.
}
\end{figure}

The free expansion of a droplet of $1.5\cdot 10^5$ 
$^6$Li atoms in the unitarity limit 
($y\simeq 0$) was investigated experimentally 
in Ref.~\cite{ohara}. 
The harmonic potential is anisotropic with 
$\lambda=\omega_z/\omega_{\rho}=0.035$. The scattering length 
for the applied magnetic field $B=910$ G is 
$a_F=-0.38\;\mu$m $=-0.14\; a_z$ \cite{converto}, 
which correspond to $y=-0.16$ at the droplet center.  
Here $a_z=[\hbar/(m \omega_z)]^{1/2}$. 
Figure~1(a) compares the observed full width half maximum (FWHM) 
of the transverse and axial size of the expanding cloud 
as a function of time \cite{ohara} with the ones obtained 
by numerical integration of the TDNLSE,  
based on the MC and EBCS equation of state. 
The TDNLSE is solved numerically 
by using a finite-difference algorithm \cite{sala0} on 
a real-space grid.

Figure 1(a) shows that the expanding gas accelerates more strongly in the
radial direction, where the confinement is tighter, than axially.
Accordingly, the cloud undergoes a shape transition: from a cigar to disk.
This is a consequence of superfluidity and interaction: a non-interacting
or a normal Fermi gas would undergo a ballistic expansion, leading
eventually to a spherical shape \cite{pitaevskii}.
In Fig.~1 the TDNLSE results are plotted as dot-dashed lines and show a
fair agreement with the experimental data.
It is important to observe that the present theory does not rely on any
fitting parameters, while the model curves shown in Ref.~\cite{ohara}
critically depend on the choice of the initial widths.
The initial profile for the TDNLSE
is obtained by running the code integrating Eq.~(6) 
in imaginary time until the confined ground-state 
is filtered out. Figure 1(b) plots the droplet 
aspect ratio showing that the theory
overestimates the experimental data. 

From the TDNLSE one can deduce Landau's hydrodynamic 
equations of superfluids at zero temperature, by 
setting $\psi({\bf r},t)= \sqrt{n({\bf r},t)} 
e^{i S({\bf r},t)}$ and ${\bf v}({\bf r},t)= (\hbar / m) 
\nabla S({\bf r},t)$, 
and neglecting the quantum-pressure term 
$(-\hbar^2 \nabla^2 \sqrt{n})/(2m\sqrt{n})$, 
that is expected to be comparably small for a 
large number $N$ of particles \cite{kim,manini,griffin}. 
These hydrodynamic equations are 
\beqa
\partial_t n + {\bf \nabla} \cdot (n {\bf v}) &=& 0 \; , 
\\
m \; \partial_t {\bf v}  + 
\nabla \left( \mu(n) + U({\bf r},t) + {1\over 2} m v^2 \right) &=& 0 \; .  
\eeqa 
In this approximation, the stationary state in the trap 
is given by the Thomas-Fermi profile 
$
n_0({\bf r}) = \mu^{-1}\left( \bar{\mu} - U({\bf r},0) \right)  
$. 
Here $\bar{\mu}$, the chemical potential of the 
inhomogeneous system, is fixed by the normalization condition 
$
N = \int d^3{\bf r} \; n_0({\bf r}) 
$.
We impose that the hydrodynamic equations 
satisfy the scaling solutions 
$ 
n({\bf r},t)= n_0\left( {x/b_1(t)} ,{y/b_2(t)} , 
{z/b_3(t)} \right)/\bar{b}(t)
$
and 
$
{\bf v}({\bf r},t) = \left( x \,{\dot b}_1(t)/b_1(t)  , 
y \,{\dot b}_2(t)/b_2(t) , z \,{\dot b}_3(t)/b_3(t) \right)
$,
where $\bar{b}(t)=\prod_{k=1}^3 b_k(t)$.
We obtain three differential equations for the 
scaling variables $b_j(t)$, with $j=1,2,3={\rho},{\rho},z$. 
These scaling differential equations depend also on the space 
vector ${\bf r}$. Only if the chemical potential satisfies a polytropic 
power law $\mu(n)=C n^\gamma$ then the space dependence drops out 
\cite{kim,minguzzi}. In our problem $\mu(n)$ is not a power law
but we expect that the dynamics can be well approximated by 
evaluating the scaling differential equations at the center 
(${\bf r}={\bf 0}$) of the cloud 
\cite{sala1}. In this case the variables $b_j(t)$ 
satisfy the local scaling equations (LSE) 
\beq 
\ddot b_j(t) + {\bar \omega}_j(t)^2 \; b_j(t) = 
\frac {\omega_j^2}{\bar{b}(t)}  \; 
{
{\partial \mu \over \partial n}\!\left(n_0({\bf 0})/\bar{b}(t)\right)   
\over 
{\partial \mu \over \partial n}\!\left(n_0({\bf 0})\right)
} 
\,.
\eeq 
The coupled ordinary differential 
equations (12) are integrated accurately and efficiently to arbitrary 
time by standard algorithms. 
We check the reliability of the LSE approach 
by comparing their numerical 
solutions to the expansion obtained by using the full TDNLSE (6), 
both within the MC equation of state (7). 
Figure 1 reports the LSE results as solid lines, 
clearly showing that the LSE are extremely accurate: 
solid lines are practically superimposed to dot-dashed lines (relative
difference $\ll 1$\%, see inset of Fig.~1(b)).
Figure 1(a) also reports the transverse and axial radii
obtained by solving the LSE with the 
chemical potential $\mu(n)$ given by the EBCS equations (8-9). 
Remarkably the mean-field EBCS results are closer 
to the experimental data than the MC results.
The two theories essentially coincide for the aspect ratio.

\begin{figure}
\centerline{\epsfig{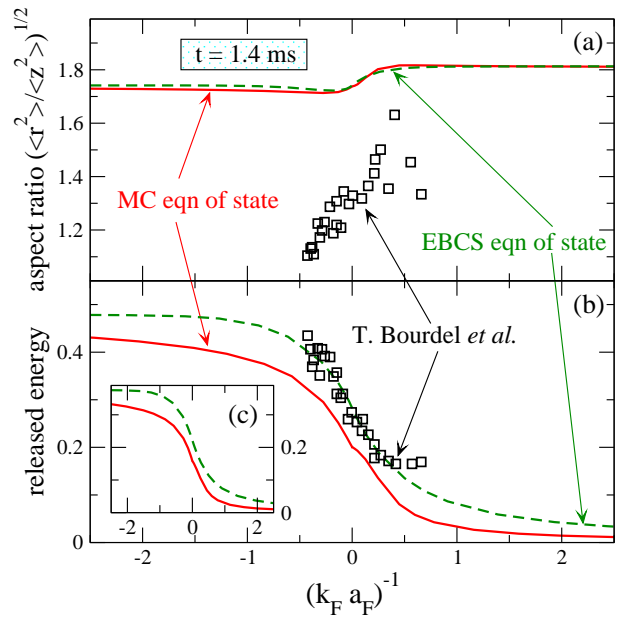}}
\caption{
(Color online). Properties of a $^6$Li cloud after 
$1.4$~ms expansion from the trap realized in Ref.~\protect\cite{bourdel}, 
of anisotrpy $\lambda=\omega_z/\omega_{\rho}=0.34$.
(a) Aspect ratio of the $^6$Li atomic cloud as a function of the inverse
interaction parameter $y=(k_Fa_F)^{-1}$.
(b) Released energy of the same cloud defined as in
Ref.~\protect\cite{bourdel} based on the rms widths of the cloud.
(c) Actual released energy of the atomic cloud.
Squares report the experimental data of Ref.~\protect\cite{bourdel}; solid
lines: LSE with the MC equation of state; dashed lines: LSE with the EBCS
equation of state.
}
\end{figure}

In Ref.~\cite{bourdel} the free expansion 
of $7\cdot 10^4$ cold $^6$Li atoms was studied 
for different values of $y=(k_Fa_F)^{-1}$ 
around the Feshbach resonance ($y=0$). Unfortunately, 
in this experiment the thermal component is not negligible 
and thus the comparison with the present $T\!=\!0$ theory is not fully 
satisfactory. Figure 2 compares the experimental data of 
Ref.~\cite{bourdel} with the LSE based on both MC and EBCS 
equation of state. 
Figure 2 shows that the aspect ratio predicted by the 
two $T\!=\!0$ theories exceeds the finite-temperature 
experimental results. This is not surprising because 
the thermal component tends to suppress the hydrodynamic 
expansion of the superflud. 
On the other hand, the released energy 
of the atomic gas is well described by 
the two $T\!=\!0$ theories, and again the mean-field theory 
seems more accurate, also probably due to the thermal component.  
For completeness, Fig.~2(c) reports the actual released energy
$\int d^3{\bf r}\, n_0({\bf r})\, {\cal E}(n_0({\bf r}))$.

\begin{figure}
\centerline{\epsfig{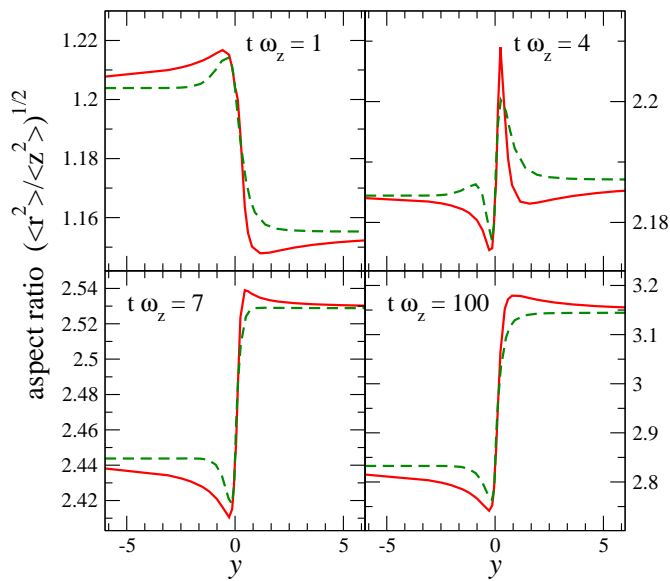}}
\caption{
(Color online). Successive frames of the 
aspect ratio of the Fermi gas as a function 
of the inverse interaction parameter $y=(k_Fa_F)^{-1}$ 
in the experimental conditions of Ref.~\protect\cite{bourdel}.  
At $t=0$ the Fermi cloud is cigar-shaped with a 
constant aspect ratio equal to the initial 
trap anisotropy $\lambda=\omega_z/\omega_{\rho}=0.34$. 
Solid lines: LSE with the MC equation of state;
dashed lines: LSE with the EBCS equation of state.} 
\end{figure}

In the two experiments of Ref.~\cite{ohara} and Ref.~\cite{bourdel} the
time evolution is sufficiently short for a full TDNLSE simulation.
It would be computationally impractical to integrate the TDNLSE for times
much longer than $\omega_H^{-1}$, where
$\omega_H=(\omega_{\rho}^2\omega_z)^{1/3}$.
As the LSE are very reliable at small and intermediate times, we use them to
investigate the time evolution of the Fermi cloud for longer times.
Figure 3 shows the aspect ratio of the expanding cloud as a function of the
inverse interaction parameter $y$ at subsequent time intervals.
At $t=0$ the aspect ratio equals the trap anisotropy $\lambda=0.34$.
According to these calculations, during the cloud expansion the aspect
ratio in the BCS regime ($y\ll -1$) is measurably different from the one of
the BEC regime ($y\gg 1$).
Thus the free expansion enables one to recognize the regime involved.
Figure 3 predicts a novel interesting effect: initially ($t\,\omega_H
\lesssim 3$) the cloud aspect ratio evolves faster in the BCS region, but
then at some intermediate time (here $t\,\omega_H \simeq 4$) the BEC side
reaches and eventually overtakes the BCS side at larger times (here
$t\omega_H \gtrsim 5$).
Of course, the detailed sequence of deformations depends on the
experimental conditions and in particular on the initial anisotropy, but
the qualitative trend of an initially faster reversal on the BCS side,
later surpassed by the BEC gas, is predicted for the expansion of any
initially cigar-shaped interacting fermionic cloud.
Similarly, starting from a disk-shaped cloud ($\lambda>1$), the aspect
ratio reduces more quickly initially on the BCS side, and later on the BEC
side.
%
%
%

\section{Discussion}

Comparison of the EBCS (dashed lines) and MC (solid lines) data shows that
beyond mean-field effects do not alter qualitatively the general trend, but
they affect the aspect ratio quantitatively, to an extent which could be
appreciated by very accurate experiments carried out at extremely low
temperature.
%
%
In particular, the mean-field curves flatten to the asymptotic values (for
$|y|\gg 1$) closer to the unitary limit than the MC ones.
Present-day experimental data, including measurements of collective
oscillation frequencies
\cite{kinast,bartenstein,stringari,combescot,minguzzi,heiselberg,kim,manini},
are equally well compatible with the EBCS mean field and the MC-based
analysis accounting for beyond mean-field effects.
New experiments could shed light on these correlation effects and verify
the predictions of the present calculations.

\acknowledgements

This work was funded in part by the EU's 6th Framework Programme through 
the NANOQUANTA Network of Excellence (NMP4-CT-2004-500198).

\end{document}